\documentclass[12pt]{raa}
\usepackage{graphicx,times}
\usepackage{natbib}
\usepackage{amssymb,amsmath}
\bibpunct{(}{)}{;}{a}{}{,}

\usepackage[pagebackref=true]{hyperref}
\hypersetup{pdftitle = The title of my PDF, pdfauthor = My name, pdfsubject= The subject, pdfkeywords = keyword1 keyword2 keyword3}
\hypersetup{colorlinks = true, linkcolor = green, anchorcolor = red, citecolor = blue, filecolor = red, pagecolor = red, urlcolor = red}

\begin{document}

   \title{On the Power-Law Distributions of X-ray Fluxes of Solar Flares Observed with the GOES
$^*$
\footnotetext{\small $*$ Supported by the National Natural Science Foundation of China.}
}

 \volnopage{ {\bf 2015} Vol.\ {\bf X} No. {\bf XX}, 000--000}
   \setcounter{page}{1}

   \author{You-ping Li, Li Feng, Ping Zhang, Siming Liu, Weiqun Gan }

   \institute{ Key Laboratory of Dark Matter and Space Astronomy, Purple Mountain Observatory,
   Chinese Academy of Sciences, 210008 Nanjing, China; {\it yplee@pmo.ac.cn; liusm@pmo.ac.cn}\\
\vs \no
   {\small Received ????; accepted ????}
}

\abstract{Power-law frequency distributions of the peak flux of solar flare X-ray emission have been studied extensively and attributed to a system of self-organized criticality (SOC). In this paper, we first show that, so long as the shape of the normalized light curve is not correlated with the peak flux,  the flux histogram of solar flares also follows a power-law distribution with the same spectral index as the power-law frequency distribution of the peak flux, which may partially explain why power-law distributions are ubiquitous in the Universe. We then show that the spectral indexes of the histograms of soft X-ray fluxes observed by GOES satellites in two different energy channels are different: the higher energy channel has a harder distribution than the lower energy channel, which challenges the universal power-law distribution predicted by SOC models and implies a very soft distribution of thermal energy content of plasmas probed by the GOES. The temperature ($T$) distribution, on the other hand, approaches a power-law distribution with an index of 2 for high values of $T$. Application of SOC models to statistical properties of solar flares needs to be revisited.
}

   \authorrunning{Y.-P. Li et al. }            
   \titlerunning{Histograms of GOES X-ray Fluxes}  
   \maketitle

%

\section{Introduction}
Power-law frequency distributions exist ubiquitously in nature, e.g., the magnitude of earthquakes \citep{Gutenberg:Richter:1954}, the frequency that a word is used in literature \citep{Zipf:1949}. More frequency distributions following a power law can be found in \citet{Clauset:etal:2009} and
\citet{Aschwanden:2011}.
For a power-law frequency distribution, the number of events $dN$ scales with the
magnitude of the event $x(>0)$ as a power-law function:
\begin{equation}
 dN=Ax^{-\delta}dx\,,
 \label{equ:power-law-diff}
\end{equation}
where the coefficient $A>0$ and the power-law index $\delta$ are constant.
Usually the distribution deviates from a power-law function towards the low end of the magnitude $x$. This deviation can be attributed either to the breakdown of the power-law scaling or to some observational bias \citep{Li:etal:2013}.

There have been quite a number of statistical works on the X-ray
emission from solar flares. The X-ray peak flux has
a power-law frequency distribution with a power-law index varying from 1.6 to 2.1 for different studies \citep[e.g.,][]{Hudson:etal:1969, Drake:1971, Shimizu:1995, Lee:etal:1995,
Feldman:etal:1997, Shimojo:Shibata:1999, Veronig:etal:2002, Yashiro:etal:2006,
Aschwanden:Freeland:2012}.
Without a background subtraction, \citet{Veronig:etal:2002} found that the peak soft X-ray (SXR) flux of flares obey a power-law distribution over three orders of magnitude from the flare GOES class C2.0 to X20. \citet{Feldman:etal:1997} divided flares observed by the GOES into different groups according
to the background level and used the background-subtracted peak flux of
flares for statistics. They found that the power law distribution can be
extended down to A1.0 class flares.
Based on the aforementioned statistical studies, \citet{Aschwanden:Freeland:2012} summarized
in their Table 2 the total
number of flares observed by the GOES, the flux range where the frequency distribution is consistent with a power-law, and the corresponding power-law indices.
They found that these observations can be explained with a fractal-diffusive avalanche model \citep{Aschwanden:2012, Du:2015}.

Although the peak flux of large flares can be easily obtained due to their high values, the value of the peak flux for small flares is always contaminated by the background emission, instrumental noise, and potential flare identification bias. The latter may be overcome by using the histogram of the flare flux, which incorporates properties of the light curve with the peak flux distribution, to study the statistics. \citet{Zhang:Liu:2015} recently showed that the characteristics of the soft X-ray light curve do not vary with the value of the peak flux. The histogram of the flare flux is therefore intimately related to the peak flux distribution.


A series of GOES satellites have taken a huge amount of SXR flux measurement of the Sun over 40 years. In this paper, the data obtained from 1981 to 2012 are used.
Moreover to reduce the effect of the selection bias, instead
of identifying individual flares, we will include all data points at an
original GOES time cadence of 3~s before 2009 and 2~s after 2009 to study the statistics of the differential histograms of the GOES fluxes, which are different from but intimately connected to the frequency distribution of the GOES peak fluxes studied before.
The GOES data reduction is presented in Section 2. The histograms are shown in Section 3. In Section 4, we explore the origin of the power law distribution of the differential histogram and its deviation from a power law towards the low value end of the flux.  Conclusions and discussions are presented in Section 5.


\section{GOES data reduction}

Since 1974, a series of GOES satellites have been put into operation and measured continuously the total soft X-ray emission flux at two wavelengths: 1-8 Å and 0.5-4 Å. The time cadence was 3 s before December 1 2009 and has been
improved to 2 s afterwards.
More information on GOES data can be found in \citet{Aschwanden:Freeland:2012} and
references therein. High temporal-resolution GOES data from 1981 to 2012 are used in this study.

\begin{figure} 
  \centering
  \includegraphics[width=\hsize]{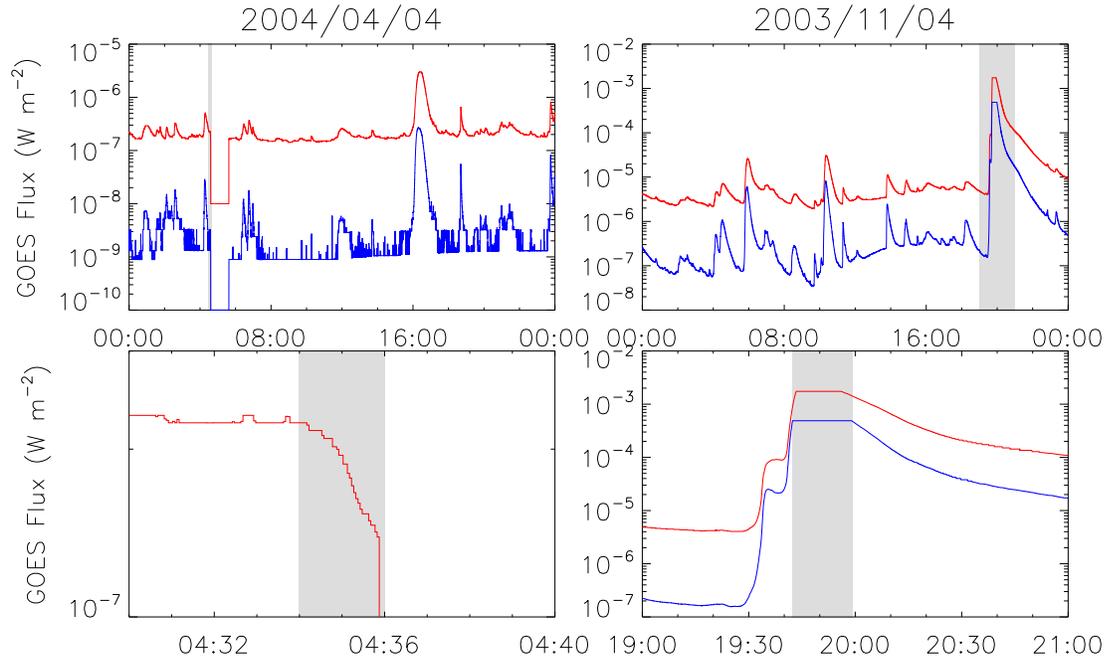}
  \caption{Two types of anomalies in the GOES data . Left panels: sudden drop of flux due
  to the Earth occultation; Right panels: instrumental flux saturation.}
  \label{fig:goes_anomolies}
\end{figure}

Before carrying out detailed investigation, one needs to treat caveats in the obtained data properly.
Two types of anomaly in the GOES data are illustrated in
Figure~\ref{fig:goes_anomolies}.
In the left panels, the sudden drop of the flux from 04:36 to 05:38~UT is due to
the entry of the satellite into the shadow of the Earth.
To remove the effect of such data on the histogram of the flux, we identify local minimums of the flux associated with the earth occultation and ignore 150 data points before and 60 data points after these minimums.

The right panels show the other type of anomaly due to instrumental saturation
\citep{Ryan:etal:2012}. 
The saturated data are marked by shaded regions
from 19:42 to 19:59~UT. This type of anomaly only affects the histogram at extremely-high flux values, which can be readily identified in the histogram.
In total, there are more than 322 million data points in our sample.

\section{Observational Results}

\begin{figure}
  \centering
  \includegraphics[width=0.9\hsize]{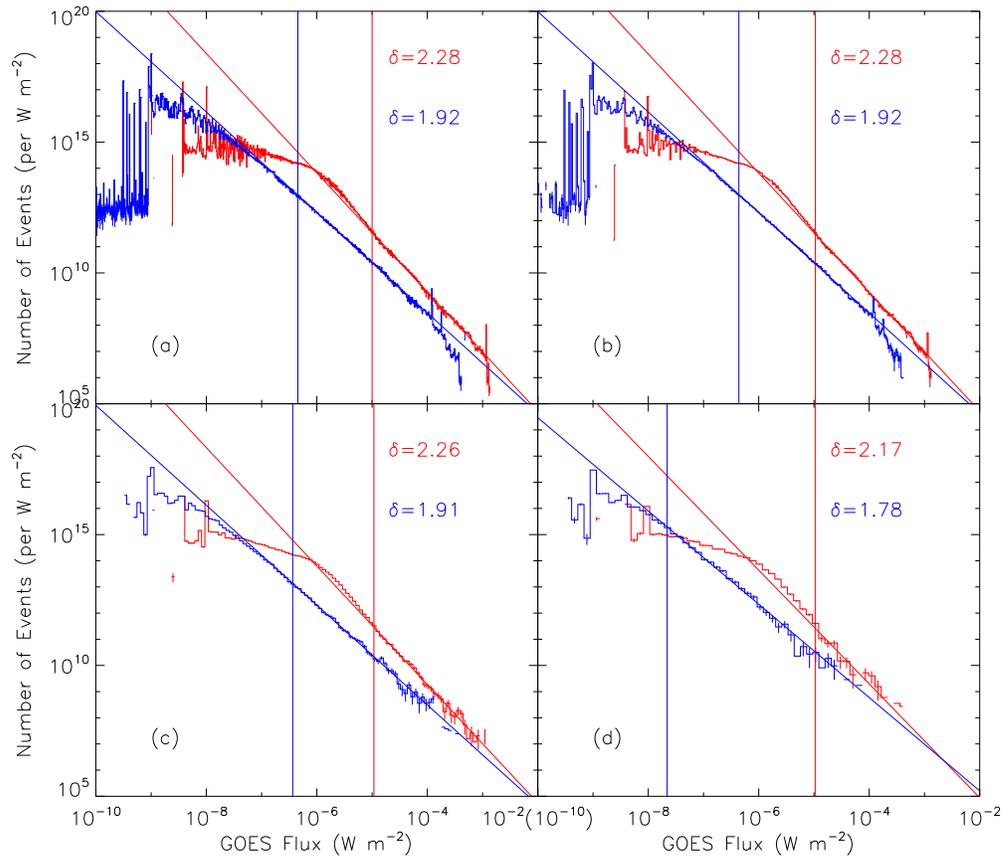}
  \caption{Differential histograms of the original data and those with coarser samplings. The red color
   is for the 1 - 8 \AA~ GOES flux, and the blue one for the 0.5 - 4 \AA~flux.
  The fittings of these histograms with a power
  law model
  \citep{Clauset:etal:2009} are indicated with straight lines with the indexes given in the corresponding figures. The
  vertical lines mark the lower cutoffs of the corresponding power laws. Panel
  (a): histograms of the full sample.
(b) histograms of the data sampled with
  a cadence of 1-minute/40-s (for a time resolution of 3s and 2s , respectively);
  (c) histograms of the data sampled with a cadence of 1-hour/40-minute; (d) histograms of the data sampled with a cadence of 1 day/16 hours.}
  \label{fig:distr_sampling_time}
\end{figure}

Panel (a) of Figure~\ref{fig:distr_sampling_time} shows the differential histograms of these fluxes in two energy channels. 
It is evident that the differential histograms of the fluxes follow a power law distribution towards the high value end of these fluxes. With the maximal-likelihood fitting procedure \citep{Crosby:etal:1993, Clauset:etal:2009, Li:etal:2012}, we fitted each histogram with a power law model. The power-law index for the lower energy band is bigger than that of the higher energy channel. The spikes at the high value end of the fluxes are caused by the flux saturation mentioned above and have been excluded in our fitting.
Panel (b-d) in Figure~\ref{fig:distr_sampling_time} are the histograms
 of a subset of the sample obtained with different sampling cadence, which have the same distribution as the one of complete sample.

\begin{figure}
  \centering
  \includegraphics[width=\hsize]{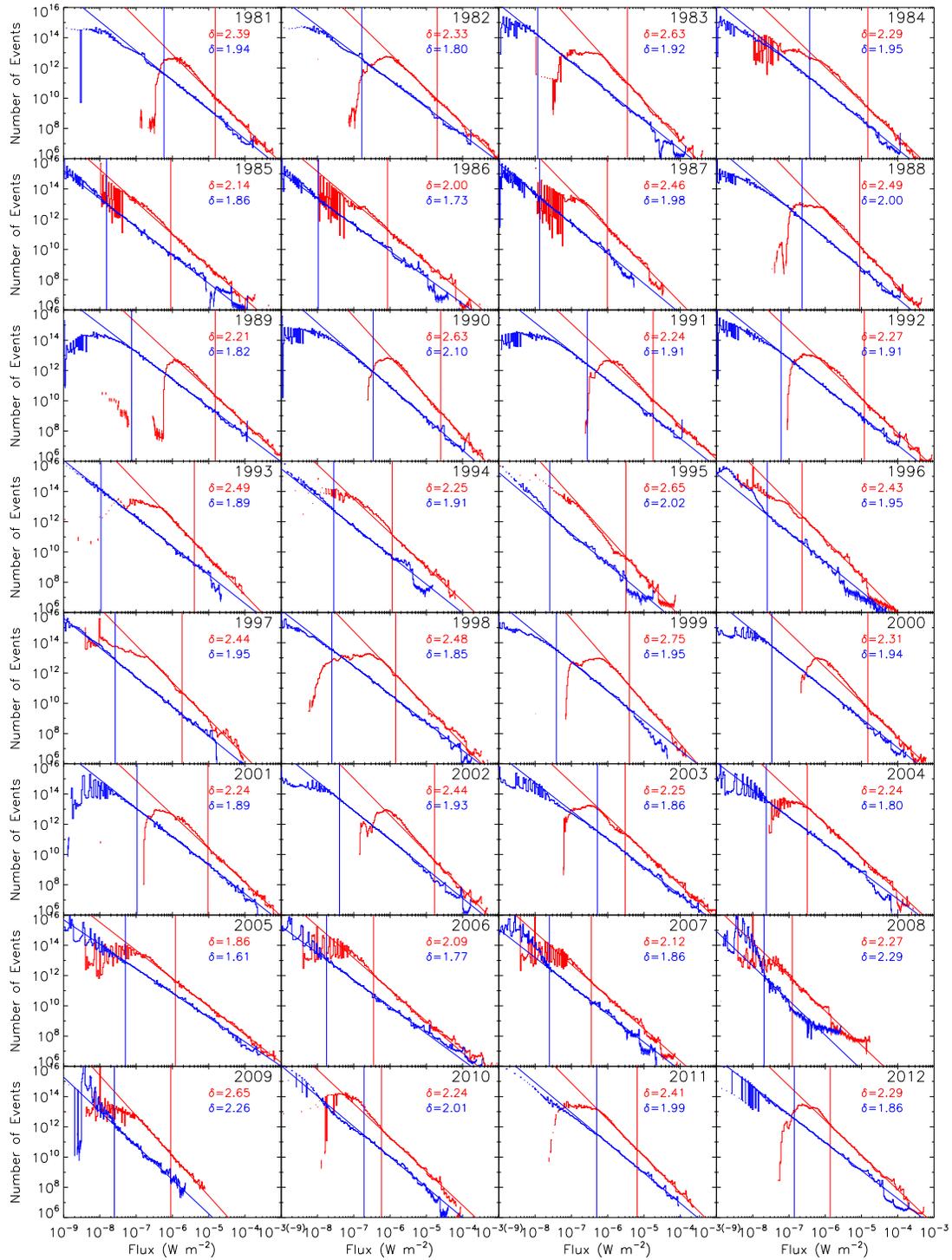}
  \caption{The histograms and their corresponding fittings in each
  year from 1981 to 2012. The year is indicated in the upper-right corner in each panel. The colors and symbols have the same meaning as
those  in Figure~\ref{fig:distr_sampling_time}.}
  \label{fig:1981_2012}
\end{figure}

To investigate the variation of the histograms in the solar cycle, different panels of Figure~\ref{fig:1981_2012} show the histograms of the flux observed in different years. 
Although the power law indexes show significant variation with the time, the index for the low energy band is always bigger than that for the high energy band.
And except in 2005 during the solar minimum, the index in the low energy band is always greater than 2, which is consistent with the frequency distribution of the peak flux \citep{Aschwanden:Freeland:2012}.

\section{Interpretation of these histograms}

Soft X-ray emission observed by the GOES is mostly produced via thermal bremsstrahlung process with the flux density given by:
\begin{equation}
F(e)\propto EM\ T^{-1/2}\exp({-e/k_{\rm B}T})\,,
\label{flux}
\end{equation}
where $T$, $EM$, and $e$ represent the plasma temperature, emission measure, and photon energy, respectively and $k_{\rm B}$ is the Boltzmann constant. The observed power-law distribution of soft X-ray fluxes in different energy bands can be used to derive the frequency distribution of  $T$ and $EM$.
From $F(e_l)^{-\delta_l} {\rm d}F(e_l) \propto F(e_h)^{-\delta_h} {\rm d}F(e_h)$, where $e_l$, $e_h$, and $\delta_l$, $\delta_h$ represent the high- and low-energy bands and the corresponding power-law indexes of their histograms, respectively, we have
\begin{equation}
 [EM\ T^{-1/2}\exp({-e_l/k_{\rm B}T})]^{1-\delta_l}\propto [EM\ T^{-1/2}\exp({-e_h/k_{\rm B}T})]^{1-\delta_h}\,,
\end{equation}
\begin{eqnarray}
EM\ & \propto & T^{1/2}\exp\{[(\delta_l-1)e_l-(\delta_h-1)e_h]/[k_{\rm B}T(\delta_l-\delta_h]\}\,,
\label{EMT}
\\
F(e_l)& \propto & \exp\{[(\delta_h-1)(e_l-e_h)]/[k_{\rm B}T(\delta_l-\delta_h]\}\,,\label{fl}\\
F(e_h)& \propto & \exp\{[(\delta_l-1)(e_l-e_h)]/[k_{\rm B}T(\delta_l-\delta_h]\}\,, \label{fh}
\end{eqnarray}
and the frequency distribution of $T$ is then given by:
\begin{equation}
D(T) \propto T^{-2} \exp\{[(\delta_l-1)(\delta_h-1)(e_h-e_l)]/[k_{\rm B}T(\delta_l-\delta_h)]\}\,.
\label{T}
\end{equation}
The frequency distribution of $EM\ T^{-1/2}$ follows a power-law with an index of
$\alpha=[\delta_l(\delta_h-1)e_h-\delta_h(\delta_l-1)e_l]/[(\delta_h-1)e_h-(\delta_l-1)e_l]\ge \delta_l$.
{\bf If $\delta_l=\delta_h$, then $\alpha=\delta_l$, and equation (\ref{flux}) implies that $T$ needs to be a constant. Therefore different spectral indexes for the histrograms of low and high channels are intimately related to the temperature distribution and the correlation between $T$ and $EM$. Equations (\ref{fl}) and (\ref{fh}) show that the harder distribution of $F(e_h)$ is caused by the greater dependence of $F(e_h)$ on $T$ than $F(e_l)$.
}

{\bf 
It is interesting to note that the temperature distribution $D(T)$ approaches a power-law distribution with an index of 2 at high values of $T$. The fast increase of the distribution toward low values of $T$ may be attributed to contribution from the background plasma no necessarily associated with individual flare events. Then in SOC models, one should associate the quantity with a universal power-law distribution with an index of 2 with intensive variable $T$ instead of fluxes or thermal energy, which are combinations of intensive and extensive variables. 

}

Equation (\ref{T}) shows that $1/T$ follows an exponential distribution with a cutoff of $[k_{\rm B}(\delta_l-\delta_h)]/(\delta_l-1)(\delta_h-1)(e_h-e_l)]$. The temperature of hot plasmas detected by the GOES therefore distributes in a relatively narraw range, which is consistent with results given by \citet{Ryan:etal:2012}. Assuming that volume $V$ of the emission region is not correlated with $T$ and density $n$, which appears to be the case for existing solar flare observations \citep{Li:etal:2012}, the themal energy of the emitting plasma is then proportional to $n V T\propto (EM V)^{1/2}T$. Considering the narrow distribution of $T$, the thermal energy of the emitting plasma therefore follows a power distribution with an index of $2\alpha-1>3$, where we have used the fact that $\alpha>\delta_l>2$.
Therefore hot plasmas with a low energy content dominates the thermal energy associated with flares \citep{Hudson:1991}.


One should emphasize that the histograms of GOES fluxes studied here are different from
frequency distributions of the flare peak flux commonly investigated. However, if
the time evolution of the flux is assumed to be independent of the peak flux:
$$
F=F_p f(t)\ ,
$$
where $0<f(t)\le1.0$ and $f(0)=1$ describe the { statistically} averaged time evolution of the normalized flare X-ray flux and
$F_p$ is the peak flux at the peak time $t=0$,
the frequency distribution of the observed flux { of a given flare} is then given by:
$$
{{\rm d}n(F)\over{\rm d} F}\equiv {\left|{\rm d}t \over T{\rm d} F\right|} = {f\over  T Ff^\prime}\ ,
$$
where $T$ is the sampling interval, $f^\prime={\rm d}f/{\rm d}t$.

{ We note that light curve of the normalized flux may vary drastically from one flare to another. However, as shown by \citet{Zhang:Liu:2015}, the statistical properties of flare light curves indeed does not vary with $F_p$. One may divide light curve of the normalized flux of all flares into several groups depending on their level of similarity with each group represented by a characteristic light curve. The $f(t)$ is then the weighted mean of these characteristic light curves.}
If $F_p$ follows a power-law distribution with an index of $\delta$ above some cutoff frequency,
the frequency distribution of $F$ for all flares is then given by:
$$
{{\rm d}N(F)\over {\rm d} F} \propto \int_F^\infty F_{p}^{-\delta}{\rm d}F_p{{\rm d}n(F)\over{\rm d} F} = F^{-\delta}\int_1^\infty f(t)^{\delta+1} {{\rm d} t(f)\over T{\rm d}f} {\rm d}\left({1\over f}\right)\,,
$$
which has the same spectral index as the distribution of $F_p$.

{\bf
The frequency distribution of the peak flux of hard X-ray emission also follows a power-law distribution.  To extrapolate the results above to higher energies, one may assume that
\begin{equation}
F(e) \propto  \exp\{[(\delta_l-1)(e_l-e)]/[k_{\rm B}T(\delta_l-\delta]\}\,, \label{f}
\end{equation}
where $\delta$ is the index of the flux distribution at $e$.
Then we have
\begin{equation}
\delta = {(\delta_l-\delta_h)(e-e_l)+(\delta_h-1)(e_h-e_l)\delta_l\over (\delta_l-\delta_h)(e-e_l)+(\delta_h-1)(e_h-e_l)}\,,
\end{equation}
which can be tested with future observations. Deviation from such a correlation between $\delta$ and $e$ will invalidate the isothermal model (\ref{flux}) for the flux.
}

\begin{figure}
  \centering
  \includegraphics[width=\hsize]{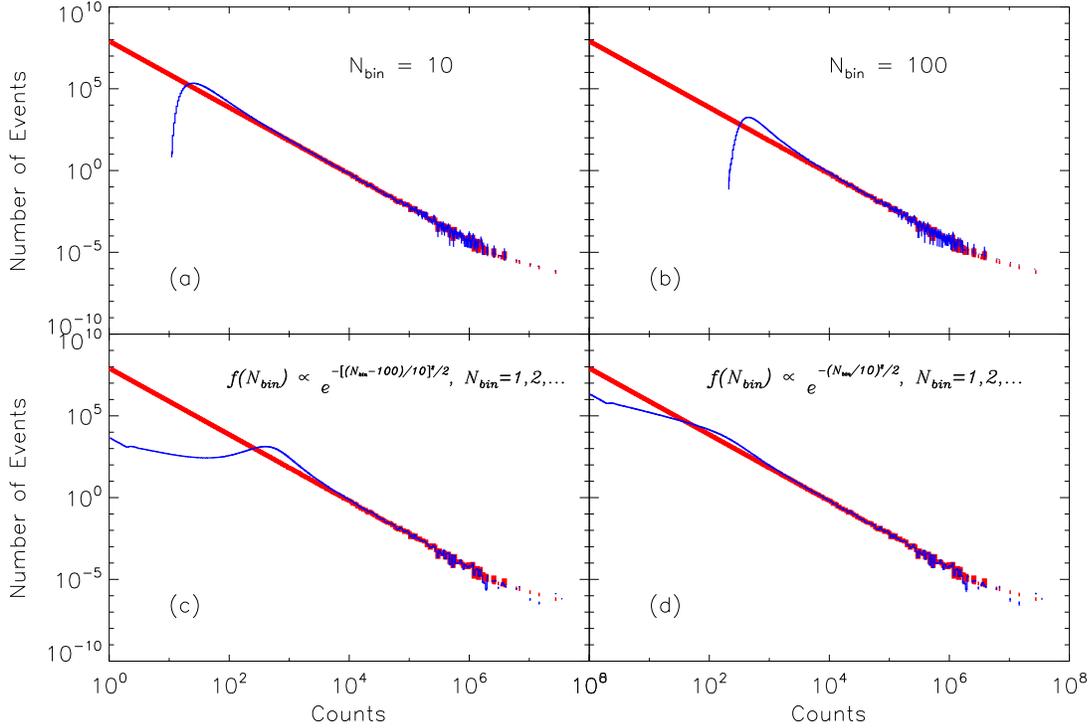}
  \caption{A demonstration of the element superposition model. The original
  frequency distribution of the counts of elements is indicated in red in each panel.
  Panel (a): randomly select $N_{bin}=10$ elements,
  and sum up their counts to form a new event point. The
  blue curve is the frequency distribution of the counts of such defined new
  event. Panel (b): Similar to the distribution in panel (a), but $N_{bin}=100$.
  Panel (c) and (d): $N_{bin}$
  follows a gaussian distribution indicated at the top of the panel. }
  \label{fig:superp_model}
\end{figure}

\begin{figure}
  \centering
  \includegraphics[width=\hsize]{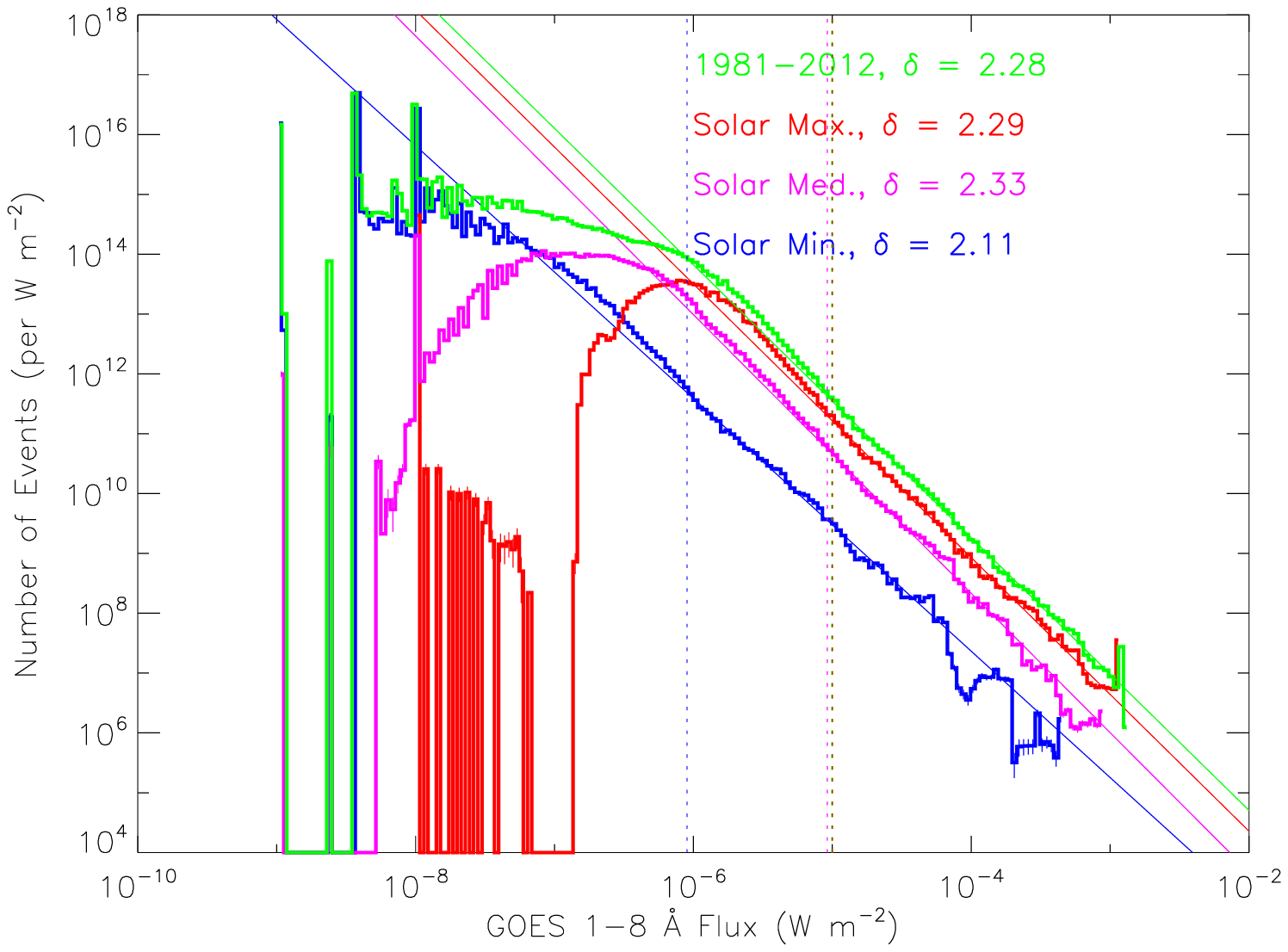}
  \caption{Histograms of GOES 1 - 8 \AA~flux at different levels
  of solar activity. The curve in red shows the distribution of the flux
  measured around the solar activity maximum.
  The curve in blue shows the result around the activity minimum. The
  one in magenta is the result derived from the flux at medium activity level.
  The green one is the distribution of the flux measured over the entire period
  from 1981 to 2012. The power-law indices of the fittings are indicated in the
  upper right corner, and vertical lines mark the flux lower cutoffs. }
  \label{fig:distr_activity}
\end{figure}


The histograms studied above deviate from a power-law distribution toward the low value end of the flux. At a given time, GOES records the soft X-ray flux coming from all features occurring in the solar disk. In fact when we check a full-disk image in SXR, e.g., an image observed by the Soft X-ray Telescope onboard YOHKOH \citep{Tsuneta:1991}, we can see active regions,
sometimes with flares superposing on it \citep[e.g.,][]{Li:etal:2012}, and bright points \cite[e.g.,][]{Shimojo:Shibata:1999, Zhang:etal:2001}, etc. Therefore, each single datum
we had in the sections above is actually the superposition of the flux from a number of
elementary phenomena. For simplicity, we call an elementary phenomenon an
element from now on.


If we assume the frequency distribution of the flux/count from these elements follows the same
power law, it would be interesting to know how the frequency distribution
produced by the superposition of a number of elements looks like. We started
from an original power-law distribution with an index of $\delta=2.03$, with
a total number of 80 million elements, and with a lower cutoff of 1. This
distribution is shown in Figure~\ref{fig:superp_model} by a thick red line. To
obtain the blue curve in panel (a), we randomly selected 10 data points along
the original power law in red, and summed their counts to form
a new data point. We repeated this process for $8\times10^7 / N_\mathrm{bin}$
times, and obtained $8\times10^7 / N_\mathrm{bin}$ samples. The frequency
distribution of the counts of this new samples is indicated by the blue curve
in Figure~\ref{fig:superp_model} (a).
One immediate finding is that after superposing 10 elements at each data point,
a ``bump'' appears in the range about 10 to 100 counts. And the drop from the 
power law distribution at the lower end is very rapid. If we increase the value of $N_\mathrm{bin}$ to 100 as
indicated in panel (b), the deviation from the power law occurs below a higher
value at about $5\times10^3$ counts, and the ``bump'' also shifts rightwards.

As the number of elements on the solar disk is not constant with time, we
may assume that $N_\mathrm{bin}$ follows a Gaussian distribution. In panels (c) and (d), we
adopt two Gaussian distributions as indicated in the upper region of both panels. When comparing the deviation part of the distribution from the power law between constant and non-constant $N_\mathrm{bin}$ cases, we find the ``bump'' regions in panels (c) and (d) are less prominent, and do not have a rapid drop toward the low end as the cases in panels (a) and (b). Interestingly, such distributions in panels (c) and (d) have similar shape to the distributions shown in Figure~\ref{fig:distr_sampling_time}.

If we attribute the ``bump'' feature in the histogram to the superposition
of elementary sources, we can expect that the histogram varies with solar activities.
At solar activity maximum, there are more elementary sources on the disk to be superposed for a measured GOES flux. While at solar activity minimum, the number of superposing elements are much lower.

To classify the 32 years from 1981 to 2012 according to the level of solar activity, we
used the flare occurring rate provided by \citet{Aschwanden:Freeland:2012} as
a criterion. Seven years had the flare occurring rate below 3000 per year are
classified to the solar minimum group. They are 1985, 1986, 1995, 1996,
2007, 2008, and 2009. Six years had the flare rate above 15000 per year belong
to the solar maximum group. They are 1989, 1990, 1991, 2000, 2001, 2002.
The solar medium activity group has the flare rate between 8000 to 13000 per
year. We found that 1983, 1992, 1993, 1998, 2004, 2011 could be included in
this group.

In Figure~\ref{fig:distr_activity}, the histograms of GOES
1 - 8~\AA~flux at different levels of solar activity are illustrated.
The curves in red, blue, and magnenta represent the histograms of the flux measured around the activity maximum, minimum, and medium, respectively.
The green curve is the histogram of the flux measured over the
entire period. As a first step to compare the results among the periods of
maximum, medium, and minimum activity, we again applied the method in
\citet{Clauset:etal:2009} to fit a power law to each curve. The best-fit
power-law indices are presented in the upper
right corner. The vertical lines mark the corresponding lower cutoffs of the
power laws. It also indicates the position below which a ``bump'' appears. We
also find that the ``bump'' at solar maximum is in a flux range about one order of
magnitude larger the ``bump'' region at solar minimum.

\begin{figure}
  \centering
  \includegraphics[width=\hsize]{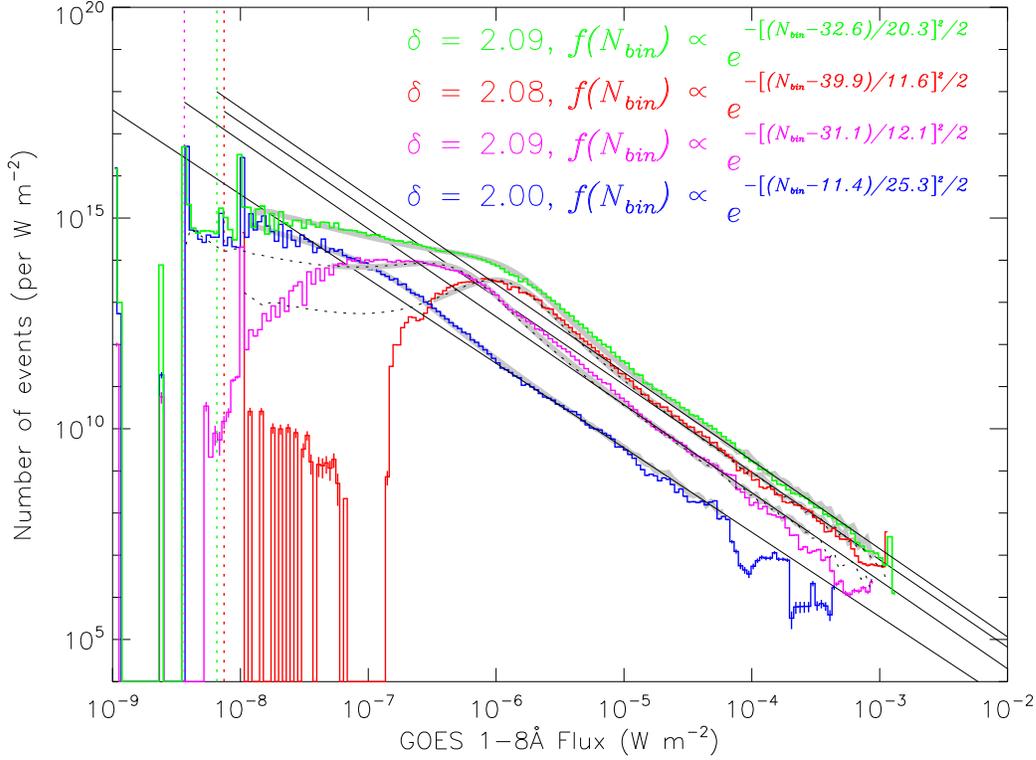}
  \caption{Fittings of the element superposition model to the histograms of GOES 1 - 8 \AA~flux at different levels of solar activity.
  Colors have the same definition as those in Figure~\ref{fig:distr_activity}.
  The solid gray lines indicate the results of the best-fit superposition models and their corresponding fitting ranges. The dashed lines represent the best-fit models to the histograms during the maximum and medium activities extrapolated down to lower flux values.
   The best-fit model parameters are marked in the upper-right region.
   The power law distributions of the elementary phenomena are shown by the black solid
  lines with their lower cutoffs marked by the vertical lines. }
  \label{fig:distri_superp_model}
\end{figure}

A more sophisticated fitting method is to use the element superposition model.
The fitting parameters include the power-law index $\delta$, the Gaussian
center and width which describes the distribution of the superposition number
$N_{bin}$, a ratio to convert counts into GOES flux, and another parameter to
convert the model frequency to the frequency of GOES flux. Therefore, there are five free parameters. In Figure~\ref{fig:distri_superp_model}
the four histograms of GOES 1 - 8 \AA~flux during the period of
minimum, medium, maximum activity, and the entire 32 years are fitted with
the element superposition model using non-linear least square method. The
best-fit power-law index and Gaussian parameters are marked in the upper right
corner.

The black solid lines are the best-fit power laws for the elementary events. They can be extended down
to the flux indicated by the vertical lines, which could be regarded as the
corresponding lower cutoffs. We can see that the lower cutoffs vary with the level of solar activity.
It implies that the distribution of these elementary events also varies with the level of solar activity.
The fittings in Figure~\ref{fig:distr_activity}
have the power-law indices of 2.11, 2.33, 2.29, and 2.28 for the period of
solar activity minimum, medium, maximum, and the entire 32 years, whereas here the fittings of
the superposition models in Figure~\ref{fig:distri_superp_model} produce harder
indices of 2.00, 2.09, 2.08, and 2.09, respectively. It is interesting to
note that the last three indices are consistent with each other,
while the distinction of the first one from the others may be attributed to
its inclusion of data from all level of the solar activity. Furthermore, the
power-law distributions of these new fittings cover much broader ranges of
GOES flux of about five orders of magnitude from about $10^{-8}$ to $10^{-3}\,\mathrm{W m^{-2}}$.

The superposition model can fit the measured histograms fairly well.
The thick gray lines in Figure~\ref{fig:distri_superp_model} represent the results of the best-fit superposition models and their corresponding fitting ranges. For the GOES flux during the entire 32 years, the superposition model is able to fit the entire range of the histogram as indicated in green color. For the GOES flux during the solar minimum, the model produces higher values than the measured
histogram at the fluxes higher than $7\times10^{-5}\,\mathrm{W\,m^{-2}}$. For the GOES flux during the solar maximum and medium, the results of the best-fit models are delineated by two black dashed lines which are extrapolated down to lower flux values. The models have higher number of events than the measured values close to the lower end.
The deviation of the measured histogram at either higher or lower flux end
from the model may be due to the selection bias when picking up the events
occurring during the period of the maximum, medium, and minimum activity
separately. The discrepancy at the lower flux end may also suggest the presence
of an X-ray background during the solar active phases, which is distinct from the elementary events proposed above.
The histogram in green color takes into account all possible events
occurring from 1981 to 2012. Therefore, it suffers the least influence of
selection bias and produces the most convincing result.

When we compare the Gaussian peak value of $N_{bin}$ in different
activity periods, we find that it is proportional to the level of
solar activity. Around solar maximum, the most probable $N_{bin}$ is about 40.
This value is larger than the one of 31.1 during the medium activity period,
and much larger than the one of 11.4 during the solar minimum. These results
could be understood straight forward, as the number of elements occurring on
the solar disk increases with the level of solar activities.

\section{Conclusions and discussions}

In this paper we have shown theoretically that if the shape of the flare light curve is not correlated with the peak flux, the differential histogram of the flare flux shares the same power-law distribution as the frequency distribution of the peak flux. Observationally, we have investigated the statistics of all usable GOES 1 - 8~\AA~ and 0.5 - 4~\AA~ flux observed from
1981 to 2012 to minimize the effect of selection bias on frequency distributions. There are two major findings in our work.

(1) The histograms of two GOES channels obey power
laws with different indices. The index of the power law for the 0.5 - 4~\AA~
GOES flux is harder than the one for the 1 - 8~\AA~ flux. And these two indices
do not change with the sampling cadence.

(2) A ``bump''-like structure is clearly seen in all the histograms
of the 1 - 8~\AA~ flux.
It
could be interpreted by the element
superposition model proposed in this paper. The original element frequency
distribution of the entire 32-year data is a power law with an index of 2.09.
This index is harder than the one derived from the fitting with the maximum
likelihood method. The best-fit parameters of superposed sources $N_\mathrm{bin}$ is correlated with
the level of solar activity.

The GOES 1 - 8~\AA~ peak flux of flares without
background subtraction has been found to follow a power law with an index
greater than 2.1, e.g., \citet{Veronig:etal:2002} obtained an index of 2.11, and
\citet{Yashiro:etal:2006} derived an index of 2.16. However, when the background
is subtracted, the 1 - 8 \AA~peak flux of flares produced a harder index a bit
below
2 \citep{Lee:etal:1995, Feldman:etal:1997, Aschwanden:Charbonneau:2002}. This
difference between with and without background subtraction can possibly be
interpreted by the element superposition model. The case with background
subtractions is equivalent to the case with less number of superposed elements,
as background subtraction would remove all other elementary sources other than
flares. Actually, the frequency distribution of the background-subtracted flare
peak flux could be linked to the upper portion of the distribution of the
elementary sources.

According to the theory of self-organized criticality (SOC) \citep{Aschwanden:2012},
\citet{Aschwanden:Freeland:2012} predicted that the peak flux of flares in SXR
has a power-law frequency distribution with an index of 2. From GOES 1 - 8~\AA~
flux measurements, the frequency distributions of background-subtracted flare
peak flux during 1975 to 2011 produced a power-law index of $1.98\pm0.11$\citep{Aschwanden:Freeland:2012}. Our
result of the power-law index of 2.09 (see Figure~\ref{fig:distri_superp_model}) is a little bit softer than the
theoretical value. The obtained power-law index in the 0.5 - 4~\AA~waveband using the maximal-likelihood fitting method (see Figure~\ref{fig:distr_sampling_time}) is about 1.92. The single event in this waveband is also a superposition of a
number of elementary sources. If we use the superposition model to derive the
original element frequency distribution, we would probably get a lower power-law
index than 1.92, which was derived using the maximum likelihood method. As the
``bump'' structure is not very pronounced in this waveband, we did not use
the superposition model for a further fitting. According to the SOC theory, the
hard X-ray (HXR) peak flux of flares have a power-law index of 1.67. The
emission from 0.5 - 4~\AA~ probably contains some contribution from HXR.
Therefore, the theoretical index might be between 1.67 to 2. Our power-law
index of $< 1.92$ is consistent with the theoretical expectation.
These results also indicate that the power-law distribution of X-ray fluxes
from solar flares involves convolution of complex physical processes over a broad
scale range and may not be simply attributed to some scaling indexes of simple mathematical models.

We have to note that the power law index of 2.09 we have derived is the
statistical result over 32 years. We have tried to minimize the bias
of event selection, as existing in flare statistics \citep{Parnell:Jupp:2000,
Aschwanden:Charbonneau:2002, Li:etal:2013}. However, if we go to the
distribution in each year as shown in Figure~\ref{fig:1981_2012}, the power-law
index for the 1 - 8~\AA~flux distribution ranges from 1.86 to 2.75, and for
the 0.5 - 4~\AA~flux, it ranges from 1.61 to 2.29. Therefore, the power-index
is quite time dependent. In particular, in year 2005 and 2008, the power
indices are very different from the values in other years.

In our simple element superposition model, the original power-law distribution
has a lower cutoff $x_0$. It is not a necessity, other forms of lower-end
deficiency can be used, such as saturation. As mentioned in the introduction,
300 samples can cover the flux in two orders of magnitude in the power-law
distribution
with an index of 2. For our 322 millions of data points, in principle they can
cover eight orders of magnitude data. However, in
Figure~\ref{fig:distr_sampling_time} the apparent power law only covers two
orders of magnitude. After removing the superposition effect, the frequency
distributions of elementary sources could be able to cover five orders of
magnitude data.

Due to the instrumental saturation of the very high GOES flux, we can not know
the exact upper limit of the flux. By extrapolating the frequency distribution
of elementary sources to the flux greater than $10^{-2}\,\mathrm{W\,m^{-2}}$
(equivalent to a X100 class flare), we find that this extremely high flux may
occur 1000 times per 32 years. As our sampling frequency is 1/3 Hz most of time,
it corresponds a time period of 3000 s, similar to the lifetime of a large
flare. That is to say, we should be able to observe a X100 class flare within
32 years. Have we observed this kind of super flares? We do not know. It may
hide in the saturated data.

\begin{acknowledgements}
This work is supported partially by the Strategic Priority Research Program,
the Emergence of Cosmological Structures, of the Chinese Academy of Sciences,
Grant No. XDB09000000, MSTC Program 2011CB811402, NSF of China under grants
11173063, 11173064, 11233008, and 11427803.
L.F. is supported by NSF of China under grants 11473070, and by the NSF of
Jiangsu Province under grants BK2012889. L.F. also
acknowledges the Youth Innovation Promotion Association, CAS, for financial
support.
\end{acknowledgements}


\begin{thebibliography}{}

\bibitem[Aschwanden \& Charbonneau(2002)]{Aschwanden:Charbonneau:2002}
Aschwanden
M.~J.,  Charbonneau P., 2002, \apjl, 566, L59

\bibitem[Aschwanden(2011)]{Aschwanden:2011} Aschwanden M.~J., 2011,
Self-organized Criticality in Astrophysics: The
Statistics of Nonlinear Processes in the Universe (Berlin: Springer)

\bibitem[Aschwanden(2012)]{Aschwanden:2012} Aschwanden M.~J., 2012, \aap, 539,
15

\bibitem[Aschwanden \& Freeland (2012)]{Aschwanden:Freeland:2012} Aschwanden
M.~J. Freeland S.~L., 2012, \apj, 754, 112

\bibitem[Crosby et al.(1993)]{Crosby:etal:1993} Crosby N.~B., Aschwanden
M.~J.,
Dennis B.~R., 1993, \solphys., 143, 275

\bibitem[Clauset et al.(2009)]{Clauset:etal:2009} Clauset A., Shalizi C.~R.,
 Newman M.~E.~J., 2009, SIAM Review, 51, 661

\bibitem[Drake(1971)]{Drake:1971} Drake J.~F., 1971, \solphys., 16, 152

\bibitem[Du(2015)]{Du:2015} Du, Z.~L., 2015, Ap\&SS, 359:4

\bibitem[Feldman et al.(1997)]{Feldman:etal:1997} Feldman U., Doschek G.~A.,
Klimchuk J.~A., 1997, \apj, 474, 511

\bibitem[Gutenberg \& Richter(1954)]{Gutenberg:Richter:1954} Gutenberg B.,
 Richter C.~F., 1954, Seismicity of the Earth and Associated
Phenomena (2nd ed.; Princeton,NJ: Princeton Univ. Press), 310

\bibitem[Han \& Liu(2013)]{2013ChA&A..37..277H} Han F.-r., Liu S.-m., 2013, Chinese Astron. Astrophys., 37, 277

\bibitem[Hudson et al.(1969)]{Hudson:etal:1969} Hudson H.~S., Peterson L.~E.,
Schwartz D.~A., 1969, \apj, 157, 389

\bibitem[Hudson(1991)]{Hudson:1991}Hudson, H.~S., 1991, \solphys., 133, 35

\bibitem[Lee et al.(1995)]{Lee:etal:1995} Lee T.~T., Petrosian V.,
McTiernan
J.~M., 1995, \apj, 418, 915


\bibitem[Li et al.(2012)]{Li:etal:2012} Li Y.~P., Gan W.~Q., Feng L.,
2012, \apj, 747, 133

\bibitem[Li et al.(2013)]{Li:etal:2013} Li Y.~P., Gan W.~Q., Feng L., Liu
S.~M., Struminsky, A., 2013, Research in Astronomy and Astrophysics, 13,
1482

\bibitem[Parnell \& Jupp(2000)]{Parnell:Jupp:2000} Parnell C.~E., Jupp P.~E.,
2000, \apj, 529, 554

\bibitem[Ryan et al.(2012)]{Ryan:etal:2012} Ryan D.~F., et al., 2012, \apjs,
202, 11

\bibitem[Shimizu(1995)]{Shimizu:1995} Shimizu T., 1995, \pasj, 47, 251

\bibitem[Shimojo \& Shibata(1999)]{Shimojo:Shibata:1999} Shimojo M.,
Shibata K., 1999, \apj, 516, 934

\bibitem[Tsuneta et al.(1991)]{Tsuneta:1991} Tsuneta S. et al., 1991,
\solphys, 136, 37

\bibitem[Veronig et al.(2002)]{Veronig:etal:2002} Veronig A.~M., Temmer M.,
Hanslmeier A., Otruba W., Messerotti M., 2002, \aap, 382, 1070

\bibitem[Yashiro et al.(2006)]{Yashiro:etal:2006} Yashiro S., Akiyama S.,
Gopalswamy N., Howard R.~A., 2006, \apjl, 650, L143

\bibitem[Zipf(1949)]{Zipf:1949} Zipf G.~K., 1949, Human Behaviour and the
Principle of Least Effort (Cambridge, MA: Addison-Wesley)

\bibitem[Zhang et al.(2001)]{Zhang:etal:2001} Zhang J., Kundu M., White
S.~M., 2001, \solphys, 88, 337


\bibitem[Zhang \& Liu (2015)]{Zhang:Liu:2015} Zhang P., Liu S. M., 2015, ChAA, 39, 330

\end{thebibliography}
\end{document}